# Tunable Moiré Bands and Strong Correlations in Small-Twist-Angle Bilayer Graphene

**Authors:** Kyounghwan Kim[1], Ashley DaSilva[2], Shengqiang Huang[3], Babak Fallahazad[1], Stefano Larentis[1], Takashi Taniguchi[4], Kenji Watanabe[4], Brian J. LeRoy[3,#], Allan H. MacDonald[2,&], and Emanuel Tutuc[1,*].

**Affiliations:**

[1]Microelectronics Research Center, The University of Texas at Austin, Austin, TX 78758, USA.

[2]Department of Physics, The University of Texas at Austin, TX 78712, USA.

[3]Department of Physics, University of Arizona, Tucson, AZ 85721, USA.

[4]National Institute for Materials Science, 1-1 Namiki Tsukuba Ibaraki 305-0044, Japan.

\* Corresponding author. E-mail : etutuc@mer.utexas.edu

& Corresponding author. E-mail: macd@physics.utexas.edu

# Corresponding author. E-mail: leroy@email.arizona.edu

**Abstract**: According to electronic structure theory, bilayer graphene is expected to have anomalous electronic properties when it has long-period moiré patterns produced by small misalignments between its individual layer honeycomb lattices. We have realized bilayer graphene moiré crystals with accurately controlled twist angles smaller than 1° and studied their properties using scanning probe microscopy and electron transport. We observe conductivity minima at charge neutrality, satellite gaps that appear at anomalous carrier densities for twist angles smaller than 1°, and tunneling densities-of-states that are strongly dependent on carrier density. These features are robust up to large transverse electric fields. In perpendicular magnetic fields, we observe the emergence of a Hofstadter butterfly in the energy spectrum, with four-fold degenerate Landau levels, and broken symmetry quantum Hall states at filling factors ± 1, 2, 3. These observations demonstrate that at small twist angles, the electronic properties of bilayer graphene moiré crystals are strongly altered by electron-electron interactions.

**Significance Statement:** Accurately controlled, very long wavelength moiré patterns are realized in small-twist-angle bilayer graphene, and studied using electron transport, and scanning probe microscopy. We observe gaps in electron transport at anomalous densities equal to ± 8 electrons per moiré crystal unit cell, at variance with electronic structure theory, and the emergence of Hofstadter butterfly in the energy spectrum in perpendicular magnetic fields. These findings open up a new avenue to create artificial crystals by manipulating the relative angle between individual layers in a heterostructure.

**Body:** Moiré patterns form when nearly identical two-dimensional crystals are overlaid with a small relative twist angle (1-4). The electronic properties of moiré crystals depend sensitively on the ratio of the interlayer hybridization strength, which is independent of twist angle, to the band energy shifts produced by momentum space rotation (5-12). In bilayer graphene, this ratio is small when twist angles exceed about 2° (10, 13), allowing moiré crystal electronic structure to be easily understood using perturbation theory (5). At smaller twist angles, electronic properties become increasingly complex. Theory (14, 15) has predicted that extremely flat bands appear at a series of magic angles, the largest of which is close to 1°. Flat bands in two-dimensional electron systems, for example the Landau level (LL) bands that appear in the presence of external magnetic fields, allow for physical properties that are dominated by electron-electron interactions, and have been friendly territory for the discovery of fundamentally new states of matter. Here we report on transport and scanning probe microscopy (SPM) studies of bilayer graphene moiré crystals with carefully controlled small-twist-angles (STA), below 1°. We find that conductivity minima emerge in transport at neutrality, and at anomalous satellite densities that correspond to ± 8 additional electrons in the moiré crystal unit cell, and that the conductivity minimum at neutrality is not weakened by a transverse electric field applied between the layers. Our observations can be explained only by strong electronic correlations.

**Methods**

Our STA bilayer graphene samples are fabricated by sequential graphene and hexagonal boron-nitride (hBN) flake pick-up steps using a hemispherical handle substrate (16) that allows an individual flake to be detached from a substrate while leaving flakes in its immediate proximity intact. In order to realize STA bilayer graphene, we start with a single graphene flake and split it into two separate sections (Fig. 1A). The separated flakes are then sequentially picked-up by an hBN flake attached to the hemispherical handle. Between the first and the second graphene flake pick-up, the substrate is rotated by a small (0.6-1.2°) angle (Fig. 1B-C) that can be controlled to

0.1° accuracy. Because the two graphene sections stem from the same crystal grain, they have crystal axes that are aligned at the onset. The substrate rotation yields a controlled twist between the two graphene layers (Fig. 1C) and forms a moiré crystal (Fig. 1D). The device fabrication is completed by encapsulating it in an hBN dielectric (17) and defining a top-gate and edge metal contacts (18) (Fig. 1E).

An example of the four-point conductance ($G$) vs. top-gate bias ($V_{TG}$) data measured in an STA bilayer graphene device at different temperatures ($T$) is shown in Fig. 1F. The data show a local conductance minimum when the carrier density ($n$) approaches zero (charge neutrality) that is similar to the minimum seen in simple gated graphene samples. In addition, two pronounced satellite conductance minima are observed at $V_{TG} = \pm 2.2$ V, corresponding to $n = \pm 2.5 \times 10^{12}$ cm$^{-2}$. All conductance minima weaken with increasing temperature, and are no longer visible at temperatures above 80 K. This striking observation departs from the density dependence of the conductance expected in either Bernal stacked (19-22) or large angle twisted bilayer graphene (23), and is instead more similar to the conductance of graphene closely aligned with an hBN substrate (2-4). As we explain below, we associate the satellite conductance minima with filling the first two bands of states produced by the moiré crystal with electrons or holes. Surprisingly, the satellite conductance minima are more pronounced than for graphene on hBN, occur at different carrier densities per moiré period, and have a temperature dependence that suggests a gap has opened for charged excitations.

**Results**

**Moiré Bloch Bands.**

In reciprocal space, the first Brillouin zone (BZ) of monolayer graphene is a hexagon. When a second layer is added with a twist angle ($\theta$) with respect to the first, the corners of the two layers BZs are displaced from each other by a wave-vector $\Delta K = 2K \sin(\theta/2)$, where $K$ is magnitude of the reciprocal lattice vector at the BZ corner at which the gap between conduction and valence bands closes in graphene. This relative rotation between the layers leads to a new (moiré) BZ with reciprocal lattice vectors that are proportional to the twist angle (Fig. 2A). In real space, the rotation is manifested by a long wavelength moiré pattern with $\lambda = \frac{a}{2\sin(\theta/2)}$ where $a$ is the graphene lattice constant. Within the moiré pattern unit cell, the local bilayer stacking configuration varies (Fig. 2B). Gradual changes between AA-like regions in which the real space hexagons lie nearly on top of each other, and alternating AB and BA (Bernal) regions in which half the lattice points of one layer are in the middle of the hexagonal plaquettes of the other are readily recognized in the generically non-crystalline structure.

A good starting point for thinking about the electronic properties of STA bilayer graphene in the non-perturbative regime is to employ the electronic structure model derived in ref. 15 that adds interlayer tunneling to individual layer Dirac models, and accounts for the structural moiré pattern illustrated in Fig. 2B through a corresponding spatial pattern in the sublattice dependence of tunneling. This model does not account for atomic relaxations in the individual layer honeycomb lattices, which are certainly present, or for electron-electron interactions. Because it has the periodicity of the moiré pattern, this model is relatively easily solved numerically by taking advantage of translational symmetry, and yields (moiré) bands in a momentum space BZ defined by the moiré pattern. Each band in the moiré BZ accommodates 4 electrons per moiré period when spin and valley degrees of freedom are included. The model's predictions for several different twist angles are summarized in Fig. 2C where we see that even for twist angles as small as $\cong$

2° (5, 10, 13), the Dirac cones (linear band crossings) of the isolated layers, centered at K and K' respectively, are still recognizable. The van Hove singularity that appears at the mid-point between K and K' (the M point) is an important feature of twisted bilayer graphene's electronic structure in the perturbative regime and produces observable features in the tunneling density-of-states (TDOS). The gap between conduction and valence bands is largest at the moiré BZ center ($\Gamma$). For twist angles smaller than $\cong 2°$ (14, 15) however, the low-energy (Fig. 2C) band structure begins to depart qualitatively from the perturbative limit, becoming highly sensitive to twist angle. The moiré bands at the first magic twist angle ($\approx 1°$) are illustrated in Fig. 2C, where we see that the lowest conduction and valence bands are extremely flat, and nearly degenerate even at the $\Gamma$ point, where some higher energy bands are also present at low energies. For twist angles below the first magic angle, the low-energy bands seem to partially recover their perturbative-regime form, except that a second pair of flat bands appears at low energy. Because the energetic widths of these moiré band can be comparable to or smaller than the characteristic energy scale for interactions between the electrons that occupy these bands, $\frac{e^2}{4\pi\varepsilon_0\varepsilon\lambda}$, physical properties in this regime might be expected to be strongly altered by correlations; $e$ is the electron charge, $\varepsilon_0$ is the vacuum dielectric permittivity, and $\varepsilon = 3$ is the hBN relative dielectric constant so $\frac{e^2}{4\pi\varepsilon_0\varepsilon\lambda} \cong 20$ meV for $\lambda \cong 20$ nm. Indeed, our observations demonstrate that this is the case.

**SPM in STA Bilayer Graphene.**

Because we expect a strong dependence of electronic properties on twist angle, it is critically important that the angle achieved by the rotation process be directly measured. To this end, we employ scanning tunneling microscopy and spectroscopy to examine the topography of the long wavelength moiré patterns and their local electronic properties (11, 24-27). The SPM samples are realized using the techniques of Fig. 1A-D but do not possess a top-gate stack. Figure 2D shows an SPM topography image of a moiré pattern with $\lambda = 20.1 \pm 0.6$ nm, corresponding to a twist angle of $\theta = 0.7 \pm 0.03°$. The AA stacked regions appear as bright spots, demonstrating that the low-energy TDOS is enhanced at these sites, consistent with theoretical studies (14). To understand the electronic properties more deeply, we perform TDOS spectroscopy measurements as a function of position. The spatially averaged results measured at back-gate voltage $V_{BG} = -31$ V applied on the doped Si substrate, such that the sample is undoped are shown in Fig. 2E (black trace). We observe two low energy TDOS peaks flanking the charge neutrality point (red solid line at the Fermi level), followed by two dips in the TDOS marked by red dashed lines. Two additional TDOS peaks occur at higher energies. Using the back-gate voltage, we tune the carrier density and find that at $V_{BG} = 2$ V the first conduction band TDOS dip aligns with the Fermi level, while the dip associated with the charge neutrality point weakens (Fig. 2E, blue trace). Using the measured moiré unit cell area $A = \frac{\sqrt{3}}{2}\lambda^2$ and the back-gate capacitance of 10.8 nF/cm², we find that the conduction band TDOS dip corresponds to a bilayer density of $7.8 \pm 0.6$ electrons per moiré unit cell, corresponding to two electrons for each spin and valley degree of freedom. This observation was reproduced in four moiré patterns with twist angles ranging between 0.7° - 1° and contrasts with observations made in the larger twist angle perturbative regime, which show dips near 4 electrons per moiré unit cell. Perhaps unsurprisingly, this observation contrasts with the density-of-states (DOS) calculated from the non-interacting electron model summarized by Fig. 2C data, which show minima at ± 4 carriers per moiré unit cell (Fig. S1). The complete evolution of the TDOS with back-gate voltage is plotted in Fig. 2F. The data show a central peak, which remains near the Fermi level for all gate voltages, splits when the sample is charge neutral ($V_{BG}$ = -31 V), and moves upward in energy relative to a more rigid background as the central peak states are occupied. The movement of the two dips associated with filling 8 electrons per moiré period are shown by the dashed lines for both the conduction and valence bands. These qualitative changes in TDOS shape with carrier density do not occur in isolated graphene sheets or at larger twist angles, and can be explained only by electron-electron interaction effects.

**Electron Transport in STA Bilayer Graphene.**

In Fig. 3A, we present the four point longitudinal resistance ($R_{xx}$) vs. $n$ measured in a series of bilayer graphene samples with twist angles $\theta < 1°$. The data show $R_{xx}$ local maxima at neutrality, along with pronounced satellite $R_{xx}$ maxima at finite $n$ values. The resistivity behavior at neutrality is similar to that in moiré crystals realized in graphene on hBN. Based on the observations in Fig. 2D-F data, the $\theta$ values were assigned assuming a carrier density of ± 8 electrons per moiré unit cell at the satellite $R_{xx}$ maxima. We note that the $\theta$ values determined using this assumption are smaller by 0.1° - 0.2° with respect to the relative rotation angle used during the graphene layers transfer. We attribute this reduction to a tendency of the two graphene monolayers to rotate towards Bernal stacking ($\theta = 0°$) during the annealing steps that follow the layer transfer, an observation verified by SPM measurements. As shown in Fig. 3B, the conductance measured in lateral moiré periods ($W/\lambda$, where $W$ is the sample width) is approximately independent of twist angle, and comparable to $e^2/h$ when $n/n_0 \cong 4$; $n/n_0 \equiv nA$ is the density per moiré unit cell, and h is the Planck constant. These data suggest that the number of quantum transport channels is proportional to $W/\lambda$.

Figure 3C shows $R_{xx}$ vs. $n$ in an STA bilayer graphene with $\theta = 0.97°$, measured at different temperatures. The data show insulating temperature dependence at $n = \pm 4.5 \times 10^{12}$ cm$^{-2}$, suggesting a gap opening at ± 8 electrons per moiré unit cell and weaker temperature dependence at charge neutrality. Figure 3D summarizes the $G$ vs. $T^{-1}$ values at charge neutrality (K), and at $n = \pm 4.5 \times 10^{12}$ cm$^{-2}$ ($\Gamma(+)$ and $\Gamma(-)$). The $\Gamma(+)$ and $\Gamma(-)$ data can be fit by assuming a combination of activated $\left( G \propto e^{-\frac{\Delta}{2kT}} \right)$, and variable range hopping $\left( G \propto e^{-\left(\frac{T_0}{T}\right)^{1/3}} \right)$ conduction processes, with an energy gap $\Delta = 15$ meV, $k$ is the Boltzmann constant, $T_0$ fitting parameter. The presence of the transport gap is qualitatively consistent with Fig. 2F data, and the gap value is comparable with the estimated interaction scale for electrons localized in one period of the moiré pattern.

A potential energy difference between the layers can dramatically change bilayer graphene electronic properties, for example giving rise to a tunable band gap in Bernal stacked bilayer graphene. To reveal the role of a potential energy difference induced by a transverse electric field ($E$) on the transport properties of STA bilayer graphene, in Fig. 3E we show a contour plot of $R_{xx}$ vs. top and back-gate biases ($V_{TG}$, $V_{BG}$) in the $\theta = 0.97°$ sample at $T = 1.5$ K. Figure 3F replots the same data as a function of the density per moiré unit cell $n/n_0$, and $E = (C_{TG} \cdot V_{TG} - C_{BG} \cdot V_{BG})/2\varepsilon_0$ where $C_{TG}$ and $C_{BG}$ are the top and back-gate capacitances. The data in Figure 3E and 3F reveal two remarkable findings. First, the $R_{xx}$ maxima are continuously present at 0, and ± 8 electrons per moiré unit cell, over a wide range of transverse $E$-fields. Most remarkable is the presence of a $R_{xx}$ maximum at charge neutrality in the entire $E$-field range, particularly since band structure calculations show that the DOS minimum at charge neutrality disappears when an on-site energy difference is applied between the layers (Fig. S2). This observation is consistent with other indications that the conductance minimum at charge neutrality is stabilized by electron-electron interactions.

Our observations are only partly understood. We associate the appearance of satellite resistance peaks at ± 8 electrons per moiré period for twist angles below the first magic angle, instead of at the ± 4 electron density expected in the perturbative regime (28, 29), with the second flat band present near the Dirac point. The appearance of an electron-electron interaction induced pseudo-gap at neutrality can be understood in terms of the expected instability of linear band crossings (Dirac bands) in two-dimensions at small Fermi velocities, and its insensitivity to a displacement field between the layers can be understood in terms of the strong hybridization between layers in low-energy bands at small twist angles (15, 30). Finally, the upward energetic shift of the central peak states relative to the background as they are filled, clearly visible in the SPM measurements, can be understood in terms of the localization of these wavefunctions near AA points in the moiré pattern.

**Magnetotransport and the Hofstadter Butterfly.**

We now turn to the magnetotransport properties of STA bilayer graphene. The energy spectrum of a two-dimensional electron system subject to a spatially periodic potential, and a perpendicular magnetic field ($B$) has a fractal structure known as the Hofstadter butterfly, characterized by two topological integers: $\nu$, representing the Hall conductivity in units of $e^2/h$, and $s$, the index of subband filling (31-34). Gaps in the energy spectrum are observed when the density per moiré unit cell and the magnetic flux per moiré unit cell ($\phi \equiv BA$) satisfy the following Diophantine equation: $\frac{n}{n_0} = \nu \frac{\phi}{\phi_0} + s$  (1); $\phi_0 = h/e$ is the quantum of magnetic flux. For $s = 0$, the integer $\nu$ reverts to the number of electrons per flux quanta, or LL filling factor. Conversely, in the limit of small $B$-fields the integer $s$ is the moiré band filling factor. A representative subset of the quantum Hall states (QHSs) that satisfy Eq. (1) are shown in Fig. 4A.

An example of $R_{xx}$ and Hall resistance ($R_{xy}$) vs. $V_{TG}$ measured at $B = 10$ T in an STA bilayer graphene sample with $\theta = 0.97°$ is shown in Fig. 4B. A contour plot of $R_{xx}$ vs. normalized density $n/n_0$ and flux $\phi/\phi_0$ is shown in Fig. 4C. The set of topological indices ($\nu$, $s$) matching the experimental data are the following. For $s = 0$, and $s = \pm 8$, QHSs are observed at $\nu = \pm 4, \pm 8, \pm 12$. For $s = 0$ developing QHSs are observed at $\nu = \pm 1, \pm 2, \pm 3$. The observation of QHSs at $\nu$ values that are multiples of four is very similar to the QHSs sequence of Bernal stacked bilayer graphene, in which orbital LLs have a four-fold, spin and valley degeneracy. On the other hand, the developing QHSs (Figs. 4B, 4C, Fig. S3) at $\nu = $ 1-3 break the spin and valley degeneracy (20, 21, 22, 35, 36). Although a full description of the ground states at these filling factors is beyond the scope of this work, these QHSs can only be stabilized by interaction, and their observation is consistent with electron-electron interactions dominating the transport properties of STA bilayer graphene at $B = 0$ T. Figure 4D shows the evolution of the QHSs as a function of $E$-field examined in the same sample by sweeping $V_{TG}$ and $V_{BG}$ at a fixed magnetic field. The data show that $R_{xx}$ at neutrality is reduced in an applied $E$-field, suggesting a weakening of the ($\nu$, $s$) = (0, 0) state. While the observation is similar to the evolution of the lowest orbital LL QHSs in Bernal stacked bilayer graphene associated with the spin-to-valley polarized transition, we note that the ($\nu$, $s$) = ($\pm$ 1, 0), ($\pm$ 2,0), ($\pm$ 3,0) remain visible in the accessible $E$-field range (Fig. S3).

In summary, we demonstrate controlled moiré crystals with long wavelengths in STA bilayer graphene, and probe the electronic properties by SPM and magneto-transport. The data reveal pseudo-gaps that open at neutrality and ± 8 electrons per moiré BZ, which are robust with respect to an applied transverse electric field, cannot be explained by electronic structure calculations, and are likely stabilized by electron-electron interaction. In high magnetic fields, we observe a Hofstadter butterfly in the energy spectrum, with subband indices of ±8, and broken symmetry states in the lowest LL.

**Acknowledgement:** This work was supported by the Semiconductor Research Corp. Nanoelectronics Research Initiative SWAN center, National Science Foundation Grants No. EECS-1610008 and EECS-1607911, the U.S. Army Research Office under Grant No. W911NF-14-1-0653, and Samsung Corp, and Welch Foundation grant TBF1473. A portion of this work was performed at the National High Magnetic Field Laboratory, which is supported by National Science Foundation Cooperative Agreement No. DMR-1157490, and the State of Florida. We thank Hema C. P. Movva for technical assistance.

**References:**

1. Xue JM, *et al.* (2011) Scanning tunnelling microscopy and spectroscopy of ultra-flat graphene on hexagonal boron nitride. *Nat Mater* 10(4):282-285.
2. Ponomarenko LA, *et al.* (2013) Cloning of Dirac fermions in graphene superlattices. *Nature* 497(7451):594-597.
3. Dean CR, *et al.* (2013) Hofstadter's butterfly and the fractal quantum Hall effect in moire superlattices. *Nature* 497(7451):598-602.
4. Hunt B, *et al.* (2013) Massive Dirac Fermions and Hofstadter Butterfly in a van der Waals Heterostructure. *Science* 340(6139):1427-1430.
5. dos Santos JMBL, Peres NMR, & Castro AH (2007) Graphene bilayer with a twist: Electronic structure. *Phys Rev Lett* 99(25):256802.
6. Mele EJ (2010) Commensuration and interlayer coherence in twisted bilayer graphene. *Phys Rev B* 81(16):161405.
7. Shallcross S, Sharma S, Kandelaki E, & Pankratov OA (2010) Electronic structure of turbostratic graphene. *Phys Rev B* 81(16):165105.
8. Yankowitz M, *et al.* (2012) Emergence of superlattice Dirac points in graphene on hexagonal boron nitride. *Nat Phys* 8(5):382-386.
9. dos Santos JMBL, Peres NMR, & Castro Neto AH (2012) Continuum model of the twisted graphene bilayer. *Phys Rev B* 86(15):155449.
10. Ohta T, *et al.* (2012) Evidence for Interlayer Coupling and Moire Periodic Potentials in Twisted Bilayer Graphene. *Phys Rev Lett* 109(18):186807.

11. Brihuega I, *et al.* (2012) Unraveling the Intrinsic and Robust Nature of van Hove Singularities in Twisted Bilayer Graphene by Scanning Tunneling Microscopy and Theoretical Analysis (2012). *Phys Rev Lett* 109(20):196802.
12. Woods CR, *et al.* (2014) Commensurate-incommensurate transition in graphene on hexagonal boron nitride. *Nat Phys* 10(6):451-456.
13. Moon P & Koshino M (2012) Energy spectrum and quantum Hall effect in twisted bilayer graphene. *Phys Rev B* 85(19):195458.
14. de Laissardiere GT, Mayou D, & Magaud L (2010) Localization of Dirac Electrons in Rotated Graphene Bilayers. *Nano Lett* 10(3):804-808.
15. Bistritzer R & MacDonald AH (2011) Moire bands in twisted double-layer graphene. *P Natl Acad Sci USA* 108(30):12233-12237.
16. Kim K, *et al.* (2016) van der Waals Heterostructures with High Accuracy Rotational Alignment. *Nano Lett* 16(3):1989-1995.
17. Dean CR, *et al.* (2010) Boron nitride substrates for high-quality graphene electronics. *Nat Nanotechnol* 5(10):722-726.
18. Wang L, *et al.* (2013) One-Dimensional Electrical Contact to a Two-Dimensional Material. *Science* 342(6158):614-617.
19. Zhang YB, *et al.* (2009) Direct observation of a widely tunable bandgap in bilayer graphene. *Nature* 459(7248):820-823.
20. Weitz RT, Allen MT, Feldman BE, Martin J, & Yacoby A (2010) Broken-Symmetry States in Doubly Gated Suspended Bilayer Graphene. *Science* 330(6005):812-816.
21. Lee K, *et al.* (2014) Chemical potential and quantum Hall ferromagnetism in bilayer graphene. *Science* 345(6192):58-61.
22. Maher P, *et al.* (2014) Tunable fractional quantum Hall phases in bilayer graphene. *Science* 345(6192):61-64.
23. Sanchez-Yamagishi JD, *et al.* (2012) Quantum Hall Effect, Screening, and Layer-Polarized Insulating States in Twisted Bilayer Graphene. *Phys Rev Lett* 108(6):076601.
24. Li GH, *et al.* (2010) Observation of Van Hove singularities in twisted graphene layers. *Nat Phys* 6(2):109-113.
25. Luican A, *et al.* (2011) Single-Layer Behavior and Its Breakdown in Twisted Graphene Layers. *Phys Rev Lett* 106(12):126802.

26. Yan W, *et al.* (2012) Angle-Dependent van Hove Singularities in a Slightly Twisted Graphene Bilayer. *Phys Rev Lett* 109(12):126801.
27. Wong D, *et al.* (2015) Local spectroscopy of moire-induced electronic structure in gate-tunable twisted bilayer graphene. *Phys Rev B* 92(15):155409.
28. Kim Y, *et al.* (2016) Charge Inversion and Topological Phase Transition at a Twist Angle Induced van Hove Singularity of Bilayer Graphene. *Nano Lett* 16(8):5053-5059.
29. Cao Y, *et al.* (2016) Superlattice-Induced Insulating States and Valley-Protected Orbits in Twisted Bilayer Graphene. *Phys Rev Lett* 117(11):116804.
30. de Laissardiere GT, Namarvar OF, Mayou D, & Magaud L (2016) Electronic properties of asymmetrically doped twisted graphene bilayers. *Phys Rev B* 93(23):235135.
31. Hofstadter DR (1976) Energy-Levels and Wave-Functions of Bloch Electrons in Rational and Irrational Magnetic-Fields. *Phys Rev B* 14(6):2239-2249.
32. Claro FH & Wannier GH (1979) Magnetic Subband Structure of Electrons in Hexagonal Lattices. *Phys Rev B* 19(12):6068-6074.
33. Macdonald AH (1983) Landau-Level Subband Structure of Electrons on a Square Lattice. *Phys Rev B* 28(12):6713-6717.
34. Bistritzer R & MacDonald AH (2011) Moire butterflies in twisted bilayer graphene. *Phys Rev B* 84(3):035440.
35. Velasco J, *et al.* (2012) Transport spectroscopy of symmetry-broken insulating states in bilayer graphene. *Nat Nanotechnol* 7(3):156-160.
36. Hunt BM, et al., (2016) Competing valley, spin, and orbital symmetry breaking in bilayer graphene. Arxiv. https://arxiv.org/abs/1607.06461.

**FIGURES**

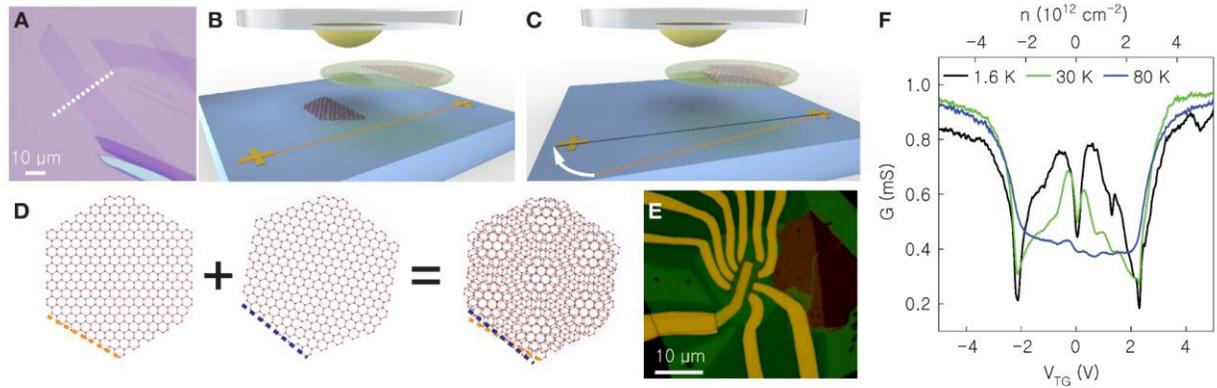

**Fig. 1. STA bilayer graphene.** (A) Optical micrograph showing a single graphene flake, subsequently split into two sections along the dotted line. (B) The first section is detached from the substrate using a hemispherical handle. (C) The second section is detached from the substrate using the same hemispherical handle. The substrate is rotated by a small angle between the two steps. Because the two flakes stem from the same graphene domain, a small twist angle is introduced between the crystal axes of the individual layers. (D) Schematic illustration of the moiré pattern formation as a result of the twist angle between the two layers. (E) Optical micrograph of an STA bilayer graphene device. (F) $G$ vs. $V_{TG}$ (bottom axis) and $n$ (top axis) measured in an STA bilayer graphene sample at different temperatures.

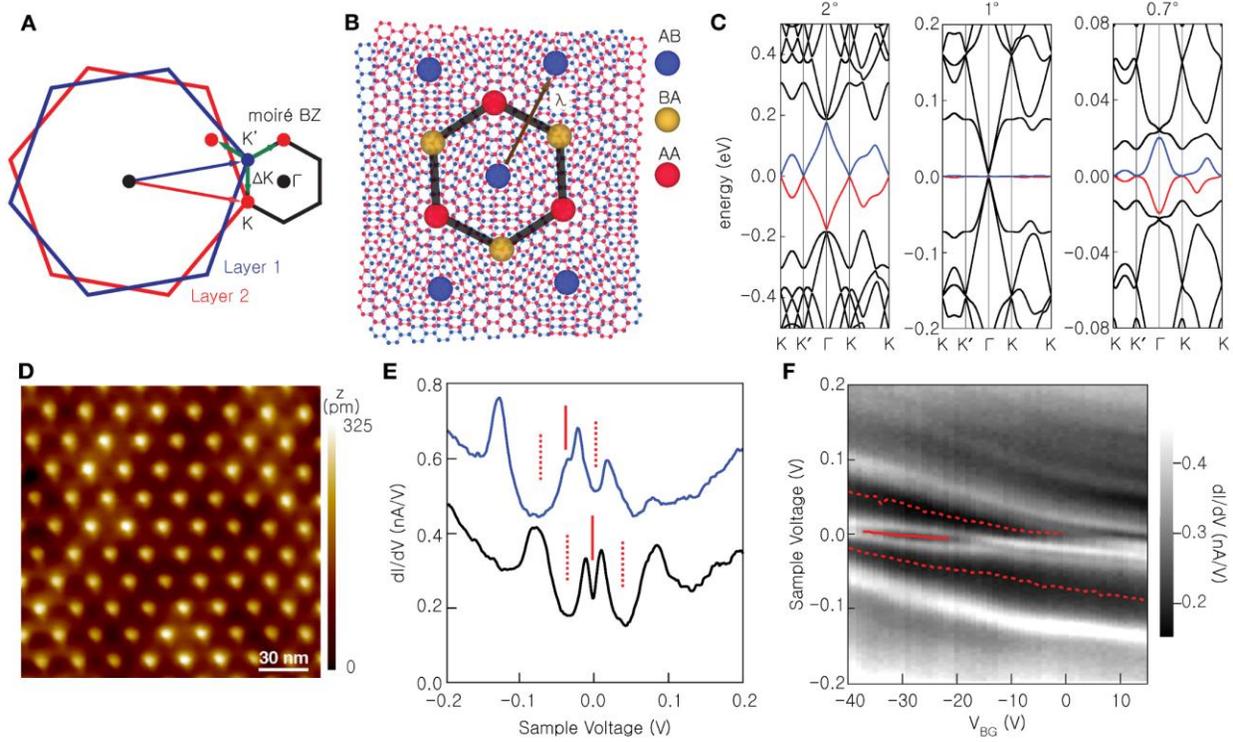

**Fig. 2. SPM mapping of the moiré pattern in a STA bilayer graphene, and DOS.** (A) The 1st BZ of each graphene layer, and of the moiré pattern. (B) Illustration of the local stacking in twisted bilayer graphene. (C) Moiré bands of twisted bilayer graphene. For twist angles $\theta$ larger than 2° most states are strongly localized in one layer, linear band crossings (Dirac points) occur at the moiré BZ corners (K, K') with velocities that are slightly suppressed relative to those of the isolated graphene, a van Hove singularity occurs at the mid-point of K and K', and a small gap opens at the center of the moiré BZ ($\Gamma$), at the extrema of the first moiré conduction and valence bands. For $\theta \approx 1°$ the first moiré bands are extremely narrow. For smaller twist angles, a second conduction and valence band are present at energies below the $\Gamma$-centered gaps. The lowest-energy bands above and below charge neutrality are shown in blue and red, respectively. (D) STM topography image showing a $\lambda = 20.1$ nm moiré pattern. The sample voltage is 0.3 V and the tunnel current is 100 pA. (E) TDOS at two different gate voltages -31 V (black) and +2 V (blue). The features corresponding to the charge neutrality point, and the secondary dips in the TDOS curves are marked by solid, and dashed lines, respectively. (F) TDOS as a function of sample voltage and gate voltage. The solid and dashed lines trace the movement of the charge neutrality point and the secondary dip in the TDOS. The data of panels (D-F) were collected at $T = 4.5$ K.

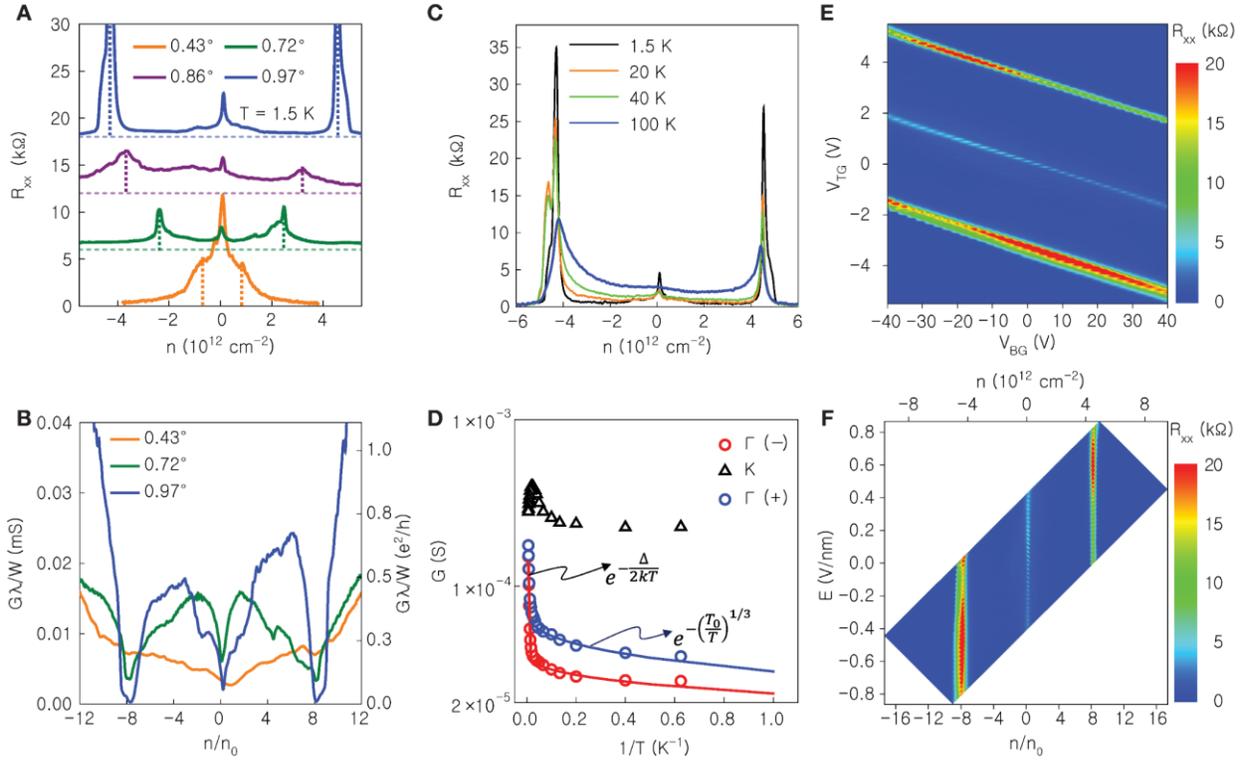

**Fig. 3. Evolution of STA bilayer graphene transport properties with twist angle, transverse E-field and temperature.** (A) $R_{xx}$ vs. $n$ measured at $T = 1.5$ K in STA bilayer graphene with different twist angles. (B) Normalized conductance ($G\lambda / W$) vs. density per moiré unit cell ($n/n_0$) in STA bilayer graphene samples with different twist angles. (C) $R_{xx}$ vs. $n$ at different temperatures measured in an STA bilayer graphene sample with $\lambda = 14.5$ nm, corresponding to $\theta = 0.97°$. (D) Arrhenius plot of $G$ measured at the K and $\Gamma$ points in the sample of panel (C). The data shows an activated dependence at elevated temperatures consistent with an energy gap, coupled with variable range hopping at low temperatures. (E) Contour plot of $R_{xx}$ vs. $V_{TG}$ and $V_{BG}$ measured at $T = 1.5$ K in the STA bilayer graphene sample of panels (C, D). The density separation between the peak at charge neutrality and the peaks at $\Gamma$ is independent of the E-field. (F) Contour plot of panel E data as a function of $n$ (top axis), $n/n_0$ (bottom axis), and transverse E-field.

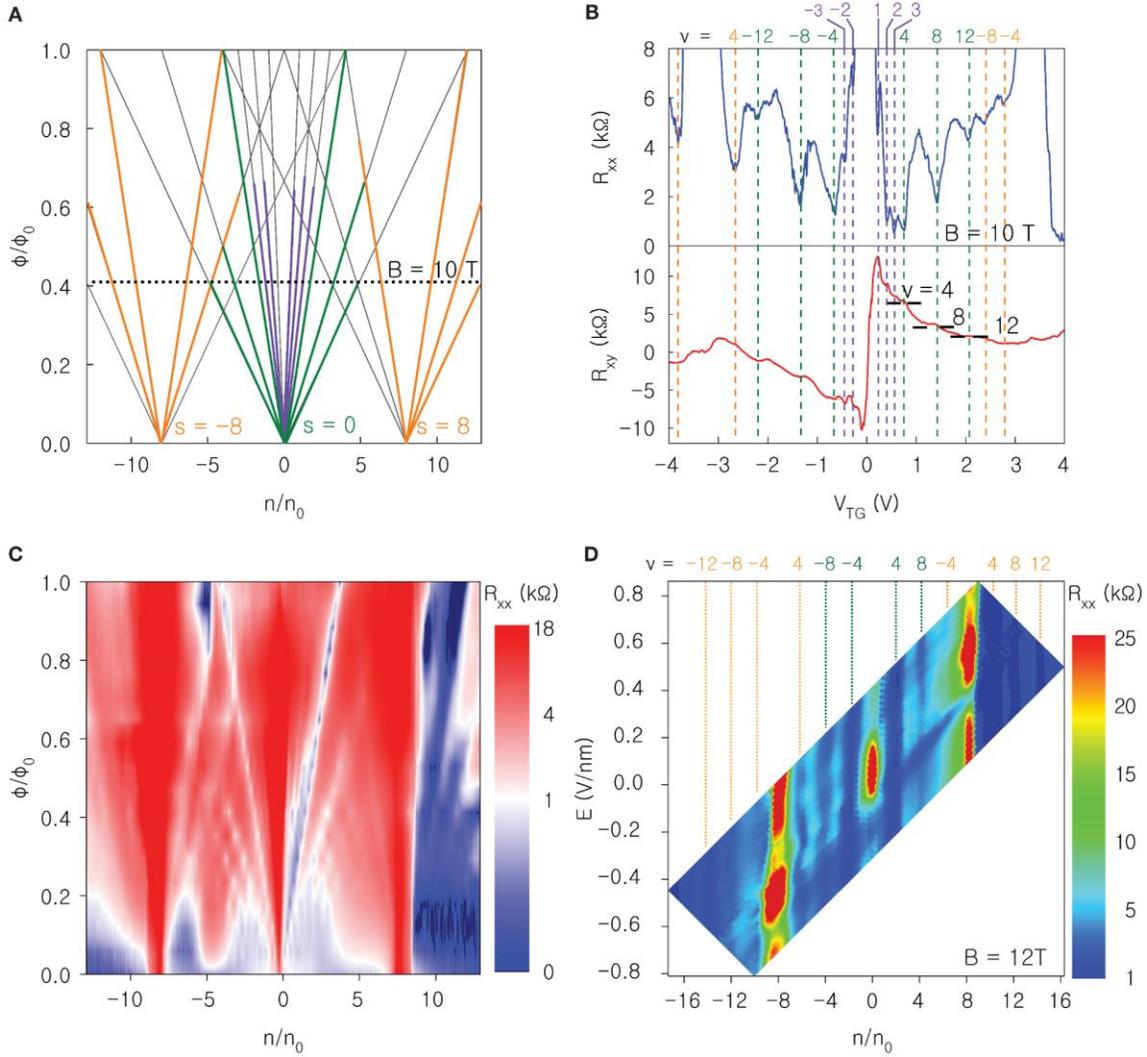

**Fig. 4. Magnetotransport properties of STA bilayer graphene.** (A) Landau level fan diagram constructed using the Diophantine equation (1). The green (orange) lines represent QHSs observed experimentally in panel (C) at $\nu = \pm 4, 8, 12$ and $s = 0$ ($s = \pm 8$). The purple lines represent broken-symmetry QHSs at $\nu = \pm 1, 2, 3$ and $s = 0$ corresponding to panel (C) data. (B) $R_{xx}$ (top panel) and $R_{xy}$ (bottom panel) vs. $V_{TG}$ measured at $B = 10$ T, and $T = 1.5$ K in an STA bilayer graphene with $\theta = 0.97°$. The $\nu$ values are marked for each QHS. The QHSs with $\nu = \pm 4, 8, 12$ and $s = 0$ ($s = \pm 8$) are marked in green (orange). The QHSs with $\nu = \pm 1, 2, 3$ are marked in purple. (C) Contour plot of $R_{xx}$ as a function of $\phi/\phi_0$ and $n/n_0$ in the same sample. The $\phi/\phi_0 = 1$ value corresponds to $B = 24.5$ T. The data are measured at $T = 1.5$ K up to $B = 14$ T, and at $T = 5$ K for $B$-fields larger than 15 T. (D) Contour plot of $R_{xx}$ vs. $n/n_0$ (bottom axis), and $E$-field, at $B = 12$ T, and $T = 1.5$ K. The $\nu$ values are marked for each QHS. The QHSs with $\nu = \pm 4, 8, 12$ and $s = 0$ ($s = \pm 8$) are marked in green (orange). Several transitions are observed as a function of the transverse $E$-field, with $R_{xx}$ at $(0, 0)$ decreasing with the applied $E$-field.

# Supplementary Information

Detailed sample fabrication process

The sample fabrication begins with micro-mechanical exfoliation of graphene and hexagonal boron-nitride (hBN) flakes on separate $SiO_2$/Si substrates. Monolayer graphene is identified using optical contrast and measuring Raman spectra 2D band full widths at half maximum, while atomic force microscope is used to determine the hBN thickness and to probe the surface topography. Once exfoliated flakes are ready, small-twist-angles (STA) bilayer graphene heterostructures are assembled using a hemispherical handle substrate dry transfer technique (*16*). The handle substrate is first coated with an adhesive polymer, poly (propylene carbonate) (PPC) or poly (methyl methacrylate) (PMMA) [poly (vinyl alcohol) PVA] for hBN encapsulated devices [layer exposed devices] which are used for transport measurement [scanning probe microscopy (SPM) measurement]. After its adhesive polymer is spun on handle substrate, the STA bilayer graphene heterostructure is fabricated using a series of angle resolved transfers, as described in the main text Fig. 1.

After the flake pick-up sequence and transfer is completed, ultra-high vacuum (UHV) annealing at 350 °C was performed to remove polymer residues. To complete the device, e-beam lithography (EBL) followed by Cr (5nm) / Au (40nm) e-beam evaporation is performed to define the top-gate of the device then, a $2^{nd}$ EBL and $CHF_3 + O_2$ plasma etching step are used to define a multi contact hall-bar shaped device. To contact the STA bilayer graphene device, we use Cr (2nm) / Pd (20nm) / Au (40nm) edge metal contacts (*18*).

The samples used for the SPM measurements are fabricated similarly to the above method, with a main difference that the STA bilayer graphene does not have a top-gate stack. Because the two layers of STA bilayer graphene have a tendency to rotate and form a Bernal stacked bilayer graphene during the annealing (*16*), to secure the flakes onto the bottom substrate and to define the metal contacts for a measurement, EBL followed by Cr (5nm) / Au (50nm) metal deposition are performed prior to the UHV anneal.

Density-of-states in twisted bilayer graphene

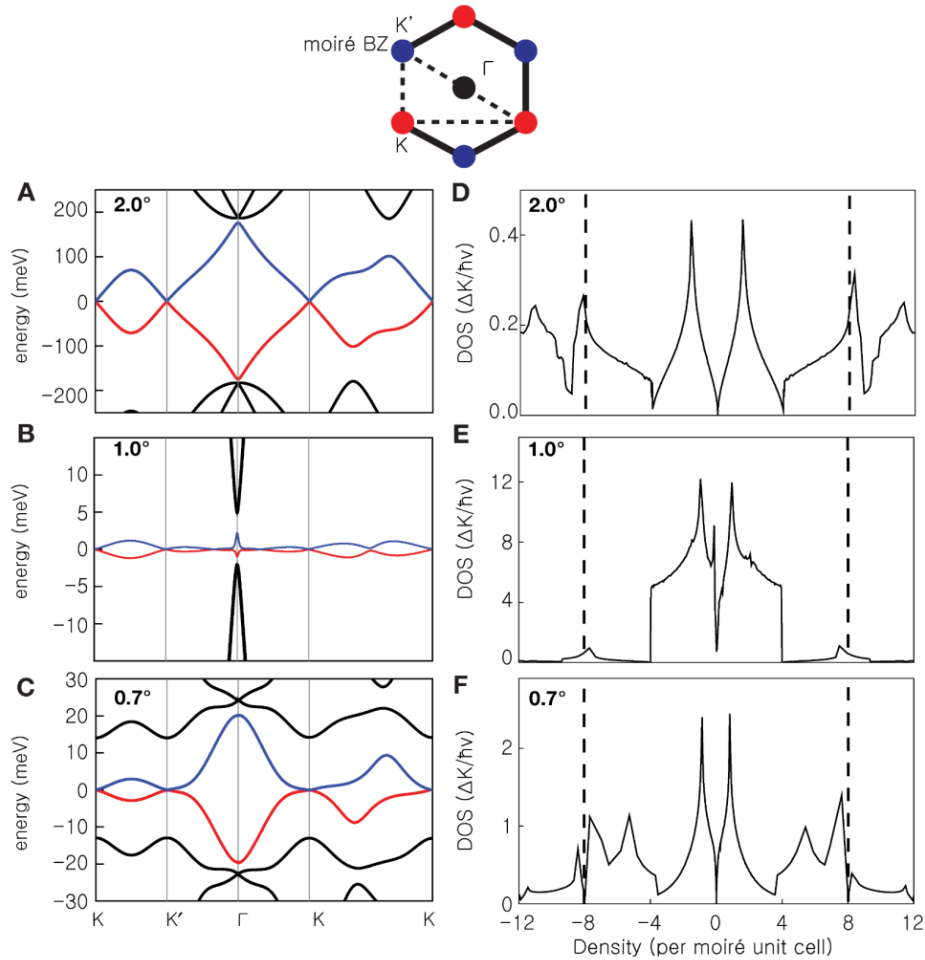

**Fig. S1. Low energy band structure (A-C) and densities-of-states (D-F) for twisted bilayer graphene.** Calculations are performed at the following twist angles: $\theta = 2°$ (panels A, D), $\theta = 1°$ (panels B, E), and $\theta = 0.7°$ (panels C, F). The path in the moiré Brillouin zone along which the band structure is calculated is shown as a dashed line in the upper cartoon. The blue and red circles represent the two inequivalent points in the moiré Brillouin zone at which the bands touch. The band structures are highly dependent on the twist angle, especially close to magic angles for which the lowest-energy bands become flat (*15*) as seen in the center panel. Independent of the twist angle, there are Dirac points at K and K' in the moiré Brillouin zone which are associated with each layer. These manifest as a zero density-of-states (DOS) at charge neutrality. The DOS has a sharp peak between 0 and ± 4 carriers per moiré unit cell, with minima close to ± 4 carriers per moiré unit cell, corresponding the filling of one band per spin and valley, for a total degeneracy of 4. Although there is a DOS minima at ± 4 carriers per moiré unit cell, experimentally there is a gap at ± 8 carriers per moiré unit cell (dashed vertical lines in DOS plots) corresponding to a filling of one band per spin, layer, and valley.

Density-of-states in twisted bilayer graphene with an on-site energy difference between the layers.

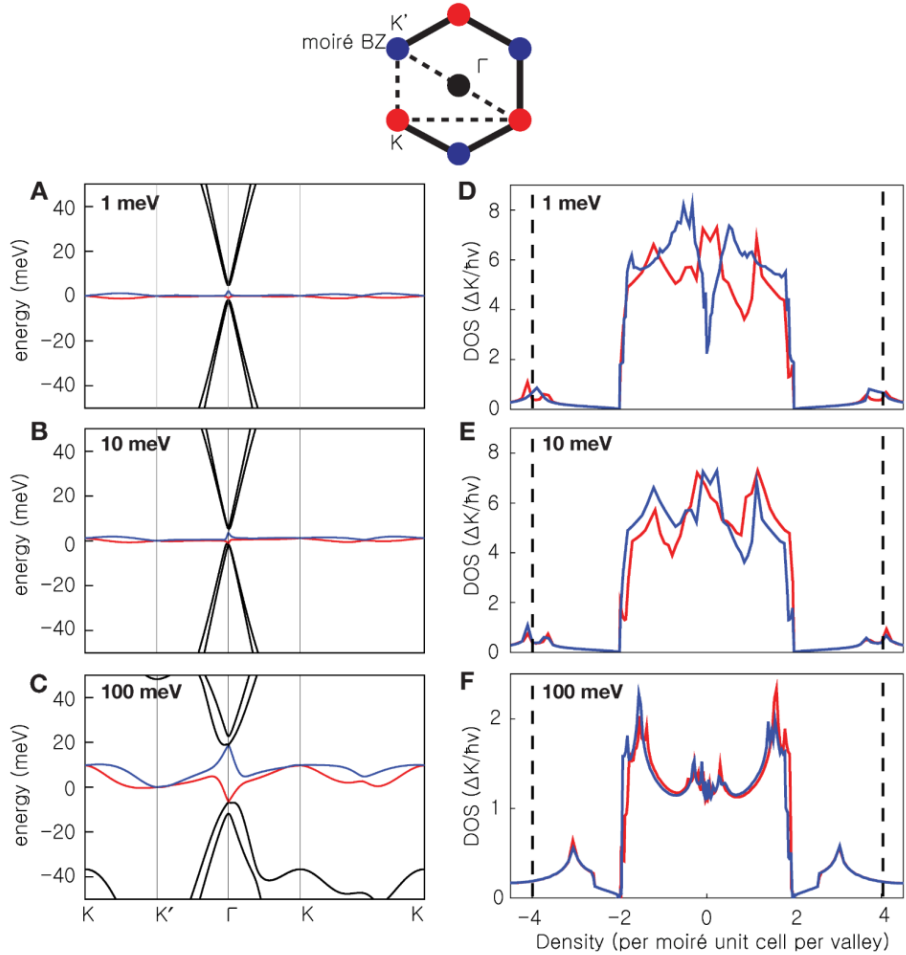

**Fig. S2. Band structure (A-C) and DOS (D-F) for twisted bilayer graphene with an on-site energy difference between the top and bottom layers, respectively.** Calculations are performed with the following + (-) on-site energy for the top (bottom) layer: 1 meV (panels A, D), 10 meV (panels B, E), and 100 meV (panels C, F). The opposite energies on each layer breaks the layer symmetry, and leads to a valley-dependent DOS, represented by blue and red lines on the DOS plots. The low-energy bands touch at K and K' points, associated with the Dirac points in each of the two layers, but their energies are shifted relative to one another. As a result, there is no longer a zero DOS at charge neutrality. However, there is still a minimum in the DOS at 4 particles per moiré unit cell (2 particles per moiré unit cell in each valley) due to the filling of 1 band per spin and valley (the densities of the two valley-dependent DOS curves must be added to get the total DOS). The dashed vertical lines in DOS plots mark the minima observed experimentally.

Broken-symmetry quantum Hall states at $\nu = \pm 1, 2, 3$

Example of developing broken symmetry $\nu = \pm 1, 2, 3$ quantum Hall states (QHSs) in an STA bilayer graphene sample with θ=0.97°, which signals a lifting of the spin and valley degeneracy in the lowest Landau level.

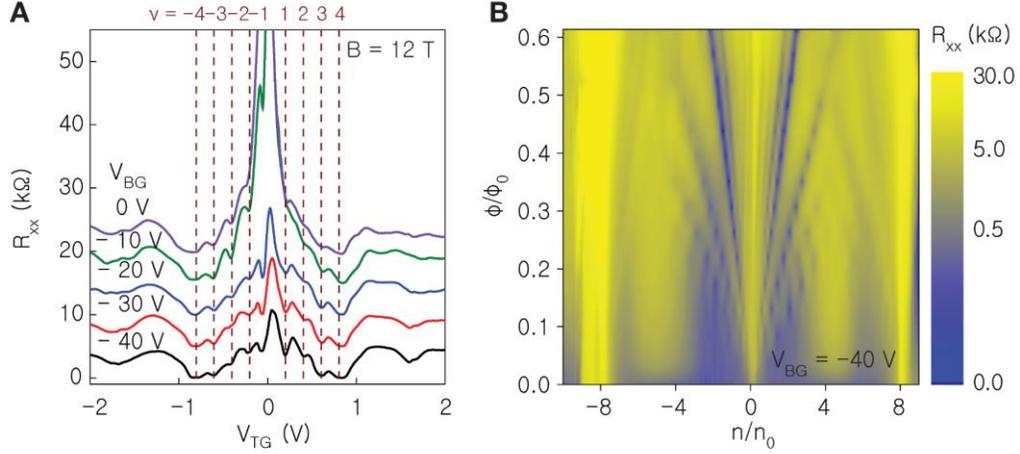

**Fig. S3. Broken-symmetry QHSs of STA bilayer graphene at $\nu = \pm 1, 2, 3$.** (A) $R_{xx}$ vs. $V_{TG}$ measured in STA bilayer graphene at $B = 12$ T, and $T = 1.5$ K, at different $V_{BG}$ from –40 V to 0V. The $R_{xx}$ vs. $V_{TG}$ traces are shifted horizontally to align the charge neutrality points at $V_{TG} = 0$ V. The $\nu = \pm 1, 2, 3, 4$ QHSs are marked with dashed lines. (B) Contour plot of $R_{xx}$ as a function of $\phi/\phi_0$ and $n/n_0$ at $V_{BG} = $ -40 V, revealing a Hofstadter butterfly with subband indices $s = \pm 8$, and broken-symmetry QHSs at $\nu = \pm 1, 2, 3$.